\documentclass[10pt]{bmc_article}

\usepackage{cite} 
\usepackage{url}  
\usepackage{ifthen}  
\usepackage{multicol}   
\usepackage[utf8]{inputenc} 
\usepackage{graphicx}
\newboolean{publ}

\newenvironment{bmcformat}{\baselineskip20pt\sloppy\setboolean{publ}{false}}{\baselineskip20pt\sloppy}

\begin{document}
\begin{bmcformat}

\title{Tamm floating electron in nanodiamond}
 
\author{Ivan A Denisov\correspondingauthor$^1$%
         \email{Ivan A Denisov\correspondingauthor - d.ivan.krsk@gmail.com}
       and
         Peter I Belobrov$^{1,2}$%
         \email{Peter I Belobrov - peter.belobrov@gmail.com}%
      }
\address{%
    \iid(1)Siberian Federal University, MOLPIT, 660074 Krasnoyarsk, Russia\\
    \iid(2)Kirensky Institute of Physics \& Institute of Biophysics SB RAS, 660036 Krasnoyarsk, Russia
}%
\maketitle

\begin{abstract}
Nanodiamond exhibits unpaired electrons in magnetization, EPR, NMR and Auger relaxation. Wave functions and eigenenergies of a bound electron in a nanodiamond crystal have been calculated. It has been proved by using quantum mechanical analysis that unpaired electrons are self-condition of a nanodiamond as a limited crystal according to Tamm theory of surface states. The surface electron floating over a nanodiamond gives paramagnetic response and stabilizes the nanoparticle at small range of size. Possibly the spin of the floating electron can be used for floating point calculation in future quantum computers on the base of nanodiamond qubits.
\end{abstract}

\ifthenelse{\boolean{publ}}{\begin{multicols}{2}}{}


\section*{Introduction}

Nowadays floating electrons in a bulk diamond are discussed as alternative to classical lead in new electronic devices \cite{Ray2011}. Laterally confined electrons above liquid helium have been demonstrated and proposed for advanced computing applications \cite{Lyon2006, Platzman1999}. Floating electrons in a nanodiamond are discussed in this paper. We are justifying here that nanodiamonds can be used as quantum dots to hold unpaired electrons in qubits.

\subsection*{Experimental evidences of surface states in nanodiamond}

Alongside with diamagnetic properties nanodiamonds demonstrate paramagnetic properties \cite{Belobrov2001, Levin2008}, which is unusual for a bulk diamond. The nature of unpaired electron spins resulting in paramagnetism is subject of great interest because unpaired electrons are not connected with d-or f-paramagnetic ions \cite{Levin2008}. $^{13}C$~NMR relaxation time underlines evidence of unpaired spins in a nanodiamond \cite{Levin2008, Fang2009} and gives approximate answer about surface localization of unpaired electrons \cite{Fang2009}. Purpose of our work is to prove, that unpaired electrons exhibited in described experiments belong to surface states of a 5nm diamond ball.

Surface localization of electronic density agrees with the PEELS scan of a single nanodiamond \cite{Peng2001, Belobrov2003}. This explains spikes (two oppositely charged layers) which depend on the size of detonation nanodiamonds \cite{Zhirnov2004} in electron emission researches. The g-factor of a unpaired electron in a nanodiamond is $2.0027$ \cite{Belobrov2001}, which is closer to free electron ($g =2.0023$) than to localized electron on atom ($\approx 1$) or NV-center. Existence and formation of NV-centres in a 5nm diamond ball is still not a solved problem \cite{Rabeau2007, Smith2009}.

\subsection*{Theory of Tamm surface states}

It is assumed, that the potential in crystals is infinite and periodic, so it  possesses translation symmetry, that is generally not true for nanocrystals. There appears always one defect --- the surface that led us to the problem of electrons on the boundary or surface states. Surface states are electronic states at surface of crystals \cite{Davison1996}. They are formed due to transition from solid to vacuum and are found only at atom layers closest to the surface. Termination of a material with surface leads to a change of electronic band structure.

Surface states were first described by I.E.~Tamm for an infinite dielectric crystal \cite{Tamm1932g}. Quantum nature of this states made them universal and identical on surface of bulk diamond and nanodiamond. I.E.~Tamm has considered diamond surface states by solving the Schröedinger equation for the Kronig-Penney potential. Tamm also predicted that electrons can move laterally on a surface like a free electron with diamond cohesion energy $0.1eV$. But still such floating electrons (surface conductivity of diamond) were not detected in bulk diamond \cite{Davison1996}. Electrons with energy $0.1eV$ correspond to de Broglie wave with $\approx 4nm$ wave length. This similarity of electron wave lengths and sizes of nanodiamods can explain both stability cases of Tamm surface states and distribution of nanodiamond size.

In the discussion of surface states, one generally distinguishes between Tamm states \cite{Tamm1932g} and Shockley states\cite{Shockley1939}. However there is no real physical distinction between these two terms. "Shockley states" term is usually used for nearly-free electron approximation for clean and ideal surfaces. "Tamm states" is mostly used for tight-binding model. But this is not consistent and Tamm and Shockley states can coexist in the same system \cite{Klos2005}.

Analysis of numerical solutions for electronic structure can give us answers to energy, electron localization in finite crystal and necessary conditions for appearing of surface states. Most important for analysis is to understand consequences of surface electron localization and effects of topological boundaries. Classical solid state theory can not to be used for nanodiamond because it is not allowed to introduce cycle boundary conditions for Bloch theorem \cite{Kittel2005} and get a good Brillouin zone for a limited crystal.

\section*{Experimental}

In this section we describe theoretical approaches and computational methods to calculations floating electrons in nanodiamond for comparison our results with experimental data.

I.E.Tamm has considered diamond surface states by solving the Schrödinger equation for the Kronig-Penney potential \cite{Tamm1932g}. An ideal solution for finding surface states in 5nm diamond ball includes the Hamiltonian for 10000 moving carbons and 60000 electrons interacting with each other including magnetic moment (spin) interactions. This many electrons problem is impossible to solve completely.

In view of small overlap between inner shell states, they can be assumed to be essentially the same as in isolated atoms. Using Born-Oppenheimer approximation nucleus positions are freezed. HOMO electrons can be imagined of as a "sea" of valence electrons moving in a crystalline lattice of nanodiamond. The mechanism looks similar to the theory of normal metals by Abricosov \cite{Abrikosov1972}, but these electrons can not be called conduction, floating or "free" electrons yet, because they interact quite strongly with ions. This strong bounding has been modelled with a deep potential holes of atoms nuclei.

So, we can look for behaviour of electrons on HOMO in a field of strongly bound electrons (such as electrons on S orbitals). Although it is impossible to calculate this field exactly, many conclusions can be drawn from symmetry properties of crystal lattice, in particular its periodicity, which an average field must possess as well \cite{Abrikosov1972}. We therefore begin with analysis of the auxiliary problem of an electron in a periodic and limited field. We consider an electron moving in an external field characterized by a potential energy $U(r)$, which is periodic and limited. This less complex calculation in one dimensional periodic potential can help to estimate Tamm states and interpret physical and chemical properties on a nanodiamond \cite{Belobrov2001, Levin2008, Fang2009, Peng2001, Belobrov2003, Zhirnov2004, Kulakova2004}.

Solutions of stationary single-electron Schrödinger equation in the framework of nearly-free electron approximation were made. Standard numerical methods were used \cite{Press1992v1} to solve and explore solutions. Software for calculation and visualization were written on Component Pascal language \cite{Mossenbock1993, Mosli1999}. Is is available at the project web page \cite{TammstatesURL} with detailed explorations of numerical methods.

\section*{Results and Discussion}

As shown in Figure \ref{fig:waveFunctions} electron density in some quantum states is localized on the surface and on energy level positioned between HOMO and LUMO (Fig. \ref{fig:energies}). This agrees with calculations of $n$-mantane ($C_{60}H_{60}$) electron structure calculations \cite{Belobrov2003}.

\begin{figure}[htb]
	\begin{center}
	\includegraphics[width=16cm]{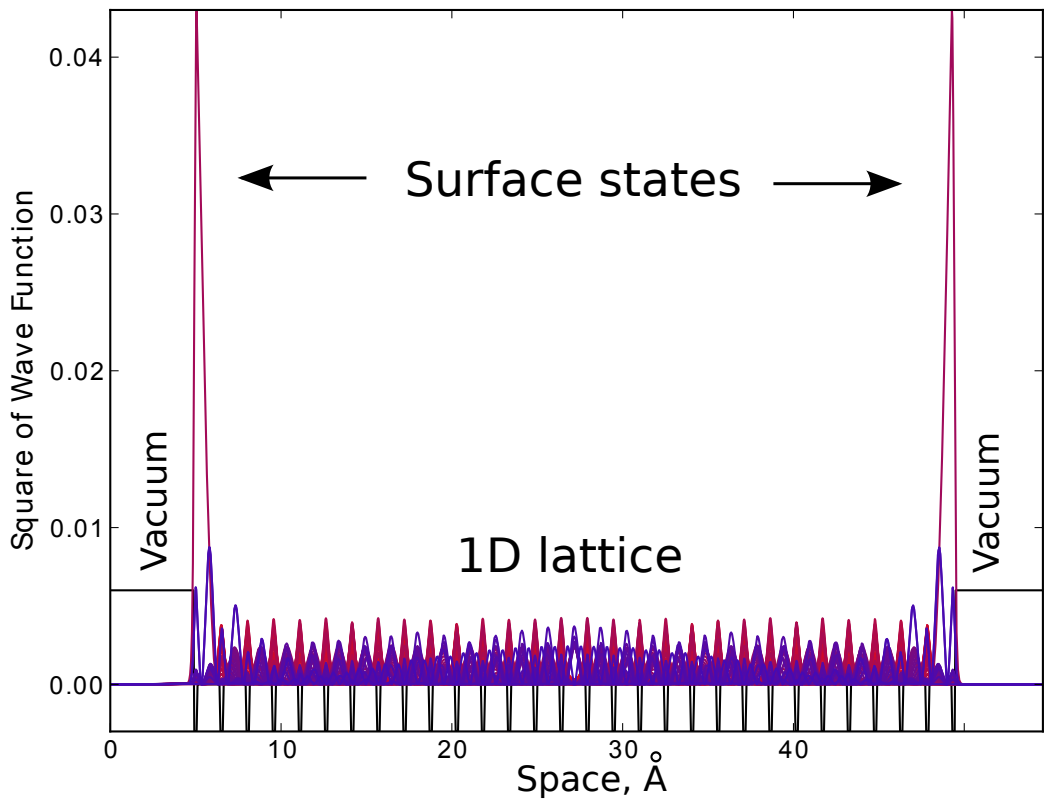}
	\end{center}
	\caption{Wave functions in limited crystal lattice}
	\label{fig:waveFunctions}
\end{figure}

The wave functions of single electron in the potential of limited lattice with different eigenenergies are shown in Figure \ref{fig:waveFunctions}. The wave functions represent valence band (red) and conductivity band (blue) for a 1D limited crystal with 30 atoms. This approximately corresponds to the 1D slice of 5nm diamond balls. Between vacuum and lattice height electronic density (bound collective electrons) is placed. Two electrons with eigenenergies between valence band and conductivity band localized on the surface between lattice and vacuum on opposite sides of the 1D crystal. In diamond conductivity band are usually empty and represented as LUMO (blue) here.

Number of states in valence band with energies including surface states is equal to number of atoms in lattice. This means that Tamm surface band is HOMO in the nanodiamond molecule, higher than the valence band (Figure \ref{fig:waveFunctions}) and always occupied with electrons from atoms of carbon (collective electrons of Tamm).

\begin{figure}[htb]
	\begin{center}
	\includegraphics[width=16cm]{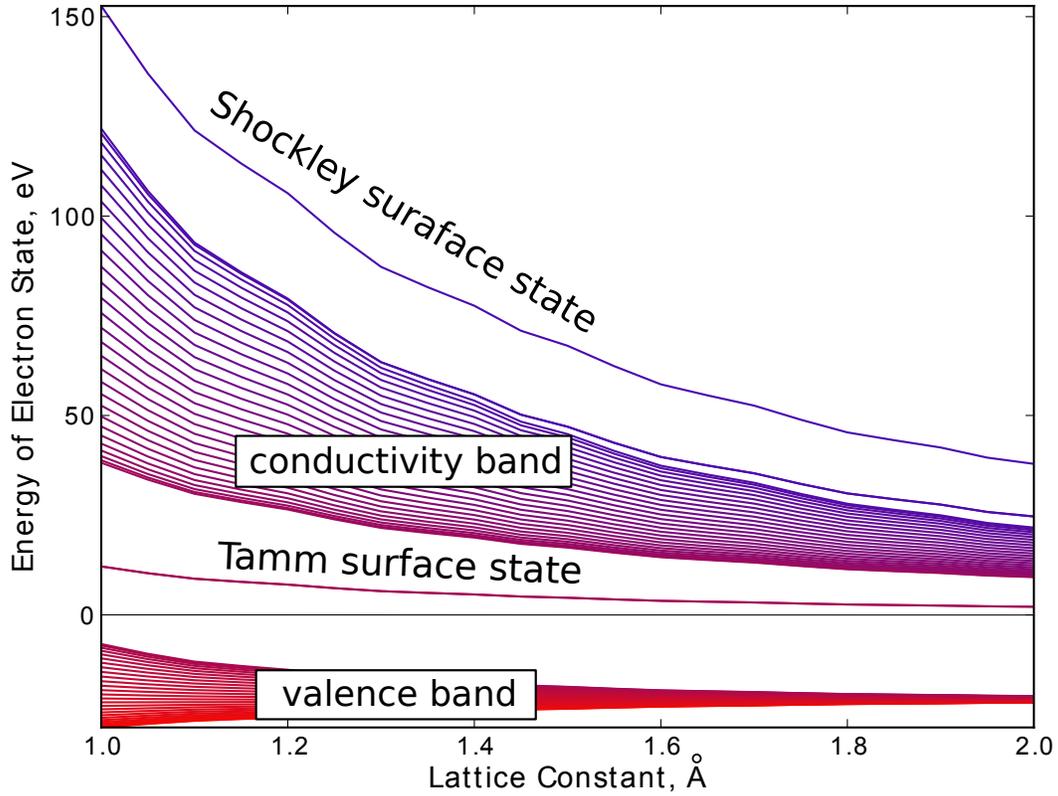}
	\end{center}
	\caption{Electron energy spectrum in limited crystal lattice}
	\label{fig:energies}
\end{figure}

The dependence of energy spectrum of electron in a limited 1D crystal on the lattice constant is presented in the Figure \ref{fig:energies}. Red lines are relevant to the electrons localized on the atoms in the crystal and represent valence band like electrons. The electrons with wave functions concentrated at the boundaries have higher energies. This level is parallel and separate to the valence band. The total amount of valence states and Tamm states is equal to the number of atoms in a 1D lattice. Lines above Tamm states present the electrons which are not localized on atoms but localized in whole crystal. This levels are empty in clear diamonds structures. Shockley states are separated from that conductivity band if they exist (for example in metals).

The one dimensional solution can be extended for discussing floating electrons in 3D nanocrystal.This can give few localized floating electrons on the surface which have radial freedom only locked by the radius. These free electrons are locally limited on the nanodiamond and appear as localized spins in different experiments.

The energy level of a Tamm floating electron is placed approximately in the middle of band gap of a nanodiamond. We can associate them with classical energy levels of vacancy in a bulk diamond.

The results and our discussion on them motivates to interpret that vacancy electron, unpaired free electron, paramagnetic electron in magnetization, free electron in EPR as well as surface states in a nanodiamond are various aspects (multyface) of a Tamm floating electron. This implicates that nanodiamond is a native quantum dot and can be the base for a qubit to hold electron.

As a result these surface electrons float easily from one atom to the next, so it is not possible to destinguish to which atom they belong to, like in the theory of normal metals \cite{Abrikosov1972}. This shared nature of the Tamm electrons is also responsible for the large cohesive energy responsible for protein adhesion of nanodiamonds and explains their specific magnetic properties.

Surface states could not be interpreted us a dangling bonds, but dangling bonds can be the source of electrons to share to whole nanoparticle to keep it charge neutral. In this case nanodiamond should be seemed by chemists like macro-molecule of new substance because the electrons are kept by each of approximately 12000 nucleus and some of them give electrons to float near the surface.

Tamm solution led to the understanding of electronic movement on a bulk single-crystal diamond surface with energy of the order of diamond cohesion energy with  0.1 eV ($800 cm^{-1}$). This energy corresponds to de Broglie wave with $\approx 4$nm which is equal to the size of thermodynamic stable nanodiamonds.

Thermodynamic stability of nanodiamond proved by fact of their fabrication in explosion method as laser ablation high purity carbon black \cite{Hu2009}. This simple comparison allowed to understand that surface quantum effects can play a bigger role in nature of nanodiamonds and make an image of de Broglie wave propagating on the angle degree of freedom inside the nanodiamond subshell wave function of surface Tamm state and somehow minimize the energy of the particle outline to the stabilizing effect.

Tamm surface states are self-consistent with all inner structure of particle. They form collective excitation of Tamm elections or quasi-particle. This Tamm quasi-particle has properties which are exhibits in Auger process, Zeeman transition, NMR relaxation and perhaps is in the cause of stability in 2 to 5nm diamond.

It can be suggested to be possible to use free electrons as spin qubits \cite{Morton2011}. Floating electrons on the nanodiamond surface can be a good alternative to the free electrons on the liquid helium surface. It would be good to organize spin-dependent optic channels, as classics from single molecule Zeeman, which allows to measure the statistics of quantum events. It is a necessary condition for formation of a nanodiamond qubits control system. Better understanding of unpaired electrons nature gives new application of floating electrons for electronics. In nearest future floating electrons in nanodiamonds can be useful for quantum computing with floating point.

Results were presented in the Nano and Giga Challenges 2011, poster \#15 \cite{Denisov2011ngs}.

\section*{Conclusions}

Nanodiamond exhibits unpaired electrons in magnetization, EPR, NMR and Auger relaxation. Wave functions and eigenenergies of a bound electron in a nanodiamond crystal have been calculated. It has been proved by using quantum mechanical analysis that unpaired electrons are self-condition of a nanodiamond as a limited crystal according to Tamm theory of surface states. The surface electron floating over a nanodiamond gives paramagnetic response and stabilizes the nanoparticle at small range of size. Possibly the spin of the floating electron can be used for floating point calculation in future quantum computers on the base of nanodiamond qubits.

\section*{Abbreviations}
NMR~---~nuclear magnetic resonance;
ND~---~nanodiamond;
NV-centres~---~nitrogen vacancy centres;
HOMO~---~highest occupied molecular orbital;
LUMO~---~lowest unoccupied molecular orbital;
EPR~---~electronic paramagnetic resonance;
PEELS~---~parallel electron energy loss spectra.

\bigskip

\section*{Competing interests}
The authors declare that they have no competing interests.

\section*{Author's contributions}
Ivan A Denisov made the concept of calculation experiments and provided them. Peter I Belobrov gave the idea of research, provided background and helped to understand results. Both authors contributed to the preparation and revision of the manuscript and approved its final version.

\section*{Acknowledgements}
  \ifthenelse{\boolean{publ}}{\small}{}
Thanks to Tobias Binder for attentive reading of the manuscript.

This research was supported by RFBR Grants 07-04-01340-a, 08-02-00259-a and (\# 09-08-98002-p-sibir-a) , ME\&S of RF Grant No. 2.2.2.2/5309 and U.S. CRDF Grant RUX0-002-KR-06/BP4M02.


\newpage
{\ifthenelse{\boolean{publ}}{\footnotesize}{\small}
 \bibliographystyle{bmc_article}  
  \bibliography{library} }     


\ifthenelse{\boolean{publ}}{\end{multicols}}{}



\section*{Figures}
  \subsection*{Figure 1 - Wave functions in limited crystal lattice}
Wave functions of a single electron in the potential of a limited lattice with different eigenenergies. Valence band (red) and conductivity band (blue) for 1D limited crystal with 30 atoms correspond to the 1D slice of a 5nm diamond ball. Between vacuum and lattice height electronic density (bound collective electrons) are placed. Two electrons with their eigenenergies between valence band and conductivity band localized in surface between lattice and vacuum on opposite sides of the 1D crystal. In diamond the conductivity band is usually empty, represented as LUMO (blue) here.

  \subsection*{Figure 2 - Electron spectrum in limited crystal lattice}
The dependence of the energy spectrum of electrons in limited 1D crystal from the lattice constant. Red lines are relevant for electrons localized on atoms in the crystal and present valence band like electrons. Electrons with wave functions concentrated at boundaries have higher energies. This lies parallel and separate to valence band. The total amount of valence states and Tamm states is equal to the number of atoms in a 1D lattice. Lines above Tamm states present the electrons which are not localized on atoms but localized in whole crystal (this levels are empty in clear diamonds structures). Shockley states are separated from that conductivity band if they exist (for example in metals).




%

\end{bmcformat}
\end{document}